\documentclass[11pt,twoside]{article}
\usepackage[pdftex]{graphicx}
\usepackage{amsmath}
\usepackage{amssymb}
\usepackage{cite}

\begin{document}
\newcommand{\pst}{\hspace*{1.5em}}

\newcommand{\rigmark}{\em Journal of Russian Laser Research}
\newcommand{\lemark}{\em Volume 30, Number 5, 2009}

%\lhead[\fancyplain{\rigmark, {\em \lemark}}{\rigmark}]{\fancyplain{\rigmark, {\em \lemark}}{\lemark}}
%\chead{}\rhead[\fancyplain{}{\lemark}]{\fancyplain{}{\rigmark}}
%\plainfootrulewidth 0.4pt
\newcommand{\be}{\begin{equation}}
\newcommand{\ee}{\end{equation}}
\newcommand{\bm}{\boldmath}
\newcommand{\ds}{\displaystyle}
\newcommand{\bea}{\begin{eqnarray}}
\newcommand{\eea}{\end{eqnarray}}
\newcommand{\ba}{\begin{array}}
\newcommand{\ea}{\end{array}}
\newcommand{\arcsinh}{\mathop{\rm arcsinh}\nolimits}
\newcommand{\arctanh}{\mathop{\rm arctanh}\nolimits}
\newcommand{\bc}{\begin{center}}
\newcommand{\ec}{\end{center}}

\thispagestyle{plain}

\label{sh}

%\lfoot[\fancyplain{\ \\[1mm] \thepage}{\ \\[1mm]\thepage}]{\fancyplain{}{}}

\begin{center} {\Large \bf
\begin{tabular}{c}
ENTROPIC AND INFORMATION INEQUALITY \\FOR NONLINEARLY TRANSFORMED TWO QUBIT X-STATES.
\end{tabular}
 } \end{center}

\bigskip

\bigskip

\begin{center} {\bf
V.I. Man'ko$^{1*}$ and R.S. Puzko$^2$
}\end{center}

\medskip

\begin{center}
{\it
$^1$P.N. Lebedev Physical Institute, Russian Academy of Science\\
Moscow, Russia 119991

\smallskip

$^2$Moscow Institute of Physics and Technology (State University)\\
Dolgoprudny, Russia 117303
}
\smallskip

$^*$Corresponding author e-mail:~~~manko~@~sci.lebedev.ru\\
\end{center}

\begin{abstract}\noindent
The entropic and information inequalities for two qubit X-states transformed by the nonlinear channels are given in explicit form. The subadditivity condition and nonnegativity of von Neumann quantum information are studied for both initial X-state and the state after action of the nonlinear channel. Partial case of Werner state is considered in detail. We shown that the von Neumann information  increases due to action of the nonlinear channel. We generalize the results obtained for Werner state in our earlier article [J. Russ. Laser Res., \textbf{35}, iss. 3 (2014)] to the case of X-state of two qubit system. We study the influence of nonlinear channel acting on  X-state of two qubit system onto von Neumann mutual information.
\end{abstract}

\medskip


\section{Introduction}
\pst
The pure quantum states are described by the vectors in the Hilbert space~\cite{1}. The mixed quantum states are described by density operators or density matrices~\cite{2,3}. The density matrices $\rho$ have the specific properties. They are Hermitian nonnegative matrices with $\mathrm {Tr}\rho=1$. The eigenvalues of the density matrices are nonnegative numbers. The linear transforms of the density matrices $\rho\to\Phi(\rho)$ which preserve the properties of the density matrices, i.e. Hermiticity, nonnegativity and trace of the matrices, are called positive maps~\cite{4}. The positive maps were studied in~\cite{4,5,6} and the maps of special kind $\Phi(\rho)=\sum_n K_n\rho K_n^+$ where $\sum_n K_n^+ K_n=1$ are called completely positive maps or quantum channels (see e.g.~\cite{7,8}). On the other hand there exist the maps of density matrices $\rho\to\Phi(\rho)$ for which the matrix $\Phi(\rho)$ has the properties of density matrix and the maps are nonlinear (see e.g.~\cite{9}). The nonlinear maps of density matrices were mentioned in~\cite{10,11}. The nonlinear map of Werner state of two qubits with rational function $\Phi(\rho)$ was studied in~\cite{12}. This positive map was called "nonlinear channel". The Werner state of two qubits is a partial case of X-state studied in~\cite{13}. Experimental study of quantum discord in the X-states was done in~\cite{14}

The aim of our work is to study the action of nonlinear quantum channels onto X-states of two qubits. We generalize our results obtained in~\cite{12} for Werner states of two qubits to the case of generic X-state of two qubits. Also we study the properties of von Neumann entropy for the X-state and states obtained by means of action of the nonlinear channel on this state. The entanglement of the nonlinearly transformed X-state of two qubits we evaluate by using concurrence~\cite{16,17}.

The paper is organized as follows. In Sec.2 properties of specific nonlinear channels of $4\times 4$ density matrix are introduced. In Sec.3 quantum tomograms~\cite{17',18',18,19,20} of the state $\Phi(\rho)$ are discussed. In Sec.4 entropic and information properties of the nonlinearly transformed X-state are studied. In Sec.5 concurrence and negativity characteristics of entanglement created by the nonlinear channels are evaluated. Conclusions and prospectives are given in Sec.6.

\section{Properties of nonlinear channels}
\pst
The nonlinear channel studied in this article is raising the density matrix to power $n$
\be
\rho_n=\frac{1}{\mathrm{Tr}\rho^n}\rho^n.
\ee
%(1)
The transformed density matrix has to be normalized to satisfy condition $\mathrm{Tr}\rho_n=1$. We consider specific state of quantum two qubit system called X-state. The density matrix of the state has the following form
\be
\rho_X=\begin{pmatrix}
a & 0 & 0 & d\\
0 & b & c & 0\\
0 & c^* & b & 0\\
d^* & 0 & 0 & a\\
\end{pmatrix},
\ee
%(2)
with parameters $a$, $b$, $c$, $d$. By varying parameters of the density matrix $\rho_X$ one can obtain states with different properties of entanglement. In particular case $a=\frac{1+p}{4}$, $b=\frac{1-p}{4}$, $c=0$, $d=\frac{p}{2}$ this density matrix describes Werner state~\cite{23} determined by parameter $p$.

The density matrix of the form (2) has the following eigenvalues
\be
\lambda_1=a+|d|, \lambda_2=b+|c|, \lambda_3=b-|c|, \lambda_4=a-|d|
\ee 
%(3) 
which are positive if $a\ge|d|$ and $b\ge|c|$. To examine the properties of entanglement we use the Peres-Horodecki criterion~\cite{21,22}. The criterion is based on positive partial transposition of density matrix $\rho\to\rho^{ppt}$. The positive partial transposed matrix (ppt-matrix) for $\rho_X$ takes form
\be
\rho_X^{ppt}=\begin{pmatrix}
a & 0 & 0 & c\\
0 & b & d & 0\\
0 & d^* & b & 0\\
c^* & 0 & 0 & a\\
\end{pmatrix}.
\ee
%(4)
The eigenvalues of the matrix $\rho_X^{ppt}$ are
\be
\lambda_{1,4}^{ppt}=a\pm|c|, \lambda_{2,3}^{ppt}=b\pm|d|.
\ee 
%(5) 
According to the criterion the state is separable if all of these eigenvalues are nonnegative. Therefore,the state is separable if the conditions 
\be\begin{matrix}
a>|c|,|d|\\
b>|c|,|d|
\end{matrix}
\ee
%(6)
are satisfied. In the domain $\left(|c|>a>|d|\right)\cup\left(b>|c|\right)$ and $\left(a>|d|\right)\cup\left(|d|>b>|c|\right)$ the density matrix $\rho_X$ corresponds to entangled state. As it can be seen from these formulas the domains do not depend on the phases of complex parameters $c$, $d$.
\begin{figure}[ht]
\bc \includegraphics[width=8.6cm]{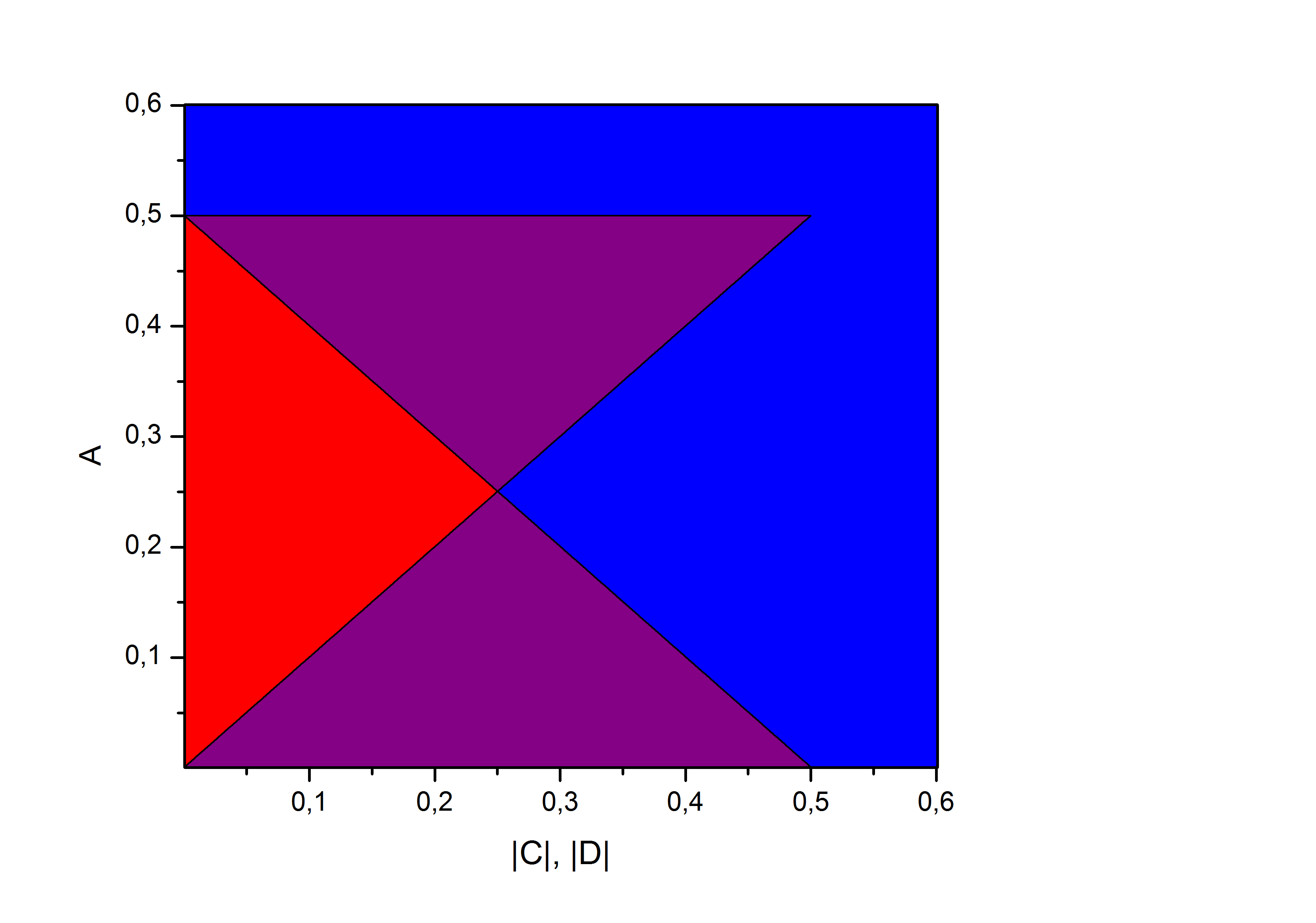}\ec
\vspace{-4mm}
\caption{Domains of parameter $|c|$ ($|d|$) where density matrix $\rho_X$ describes separable (red color) and entangled (purple color). The blue domain corresponds to set of parameters, where the matrix $\rho_X$ is not the density matrix}
\end{figure}

\begin{figure}[ht]
\begin{minipage}[h]{0.47\linewidth}
\center{\includegraphics[width=1\linewidth]{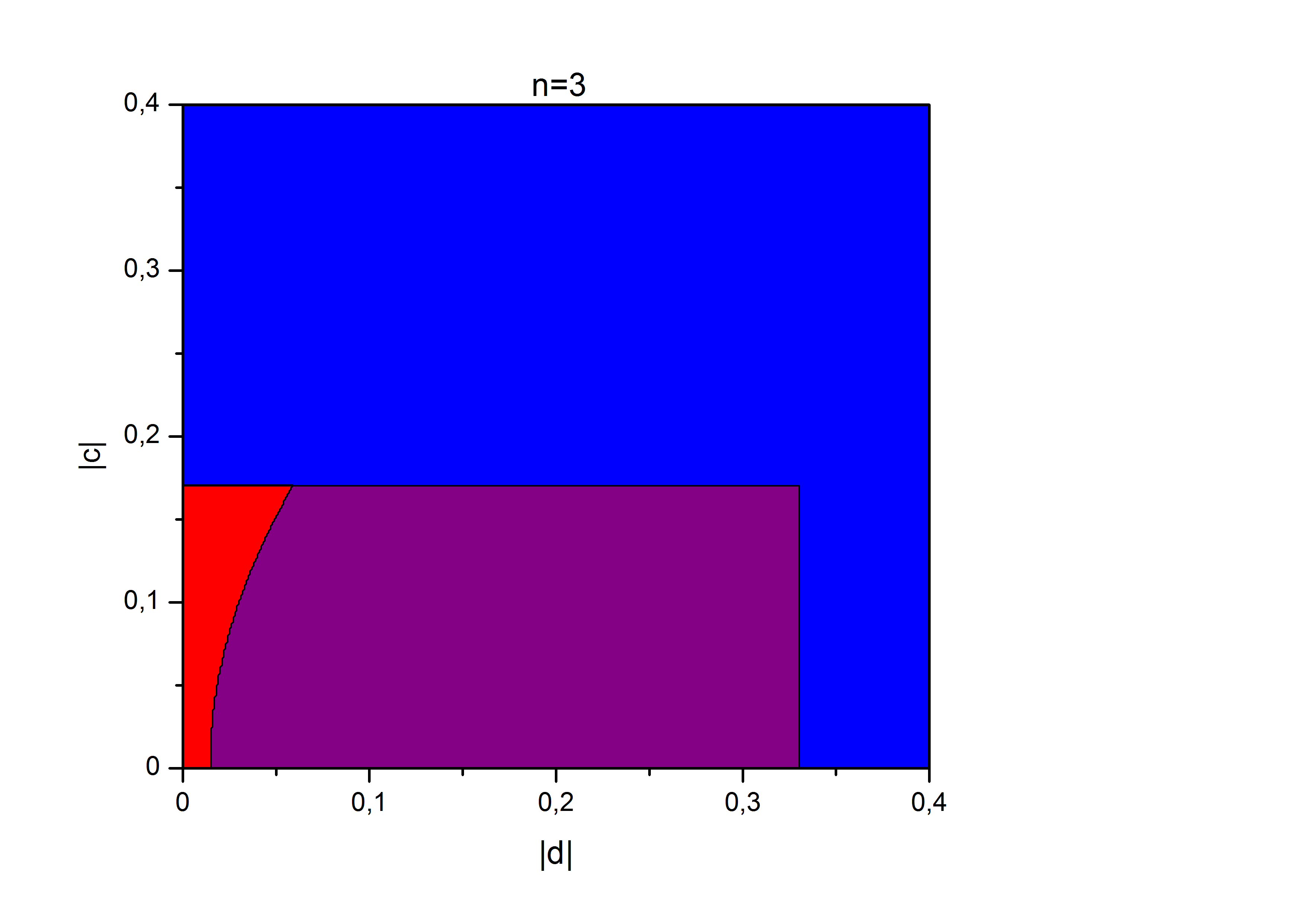}} a)\\
\end{minipage}
\hfill
\begin{minipage}[h]{0.47\linewidth}
\center{\includegraphics[width=1\linewidth]{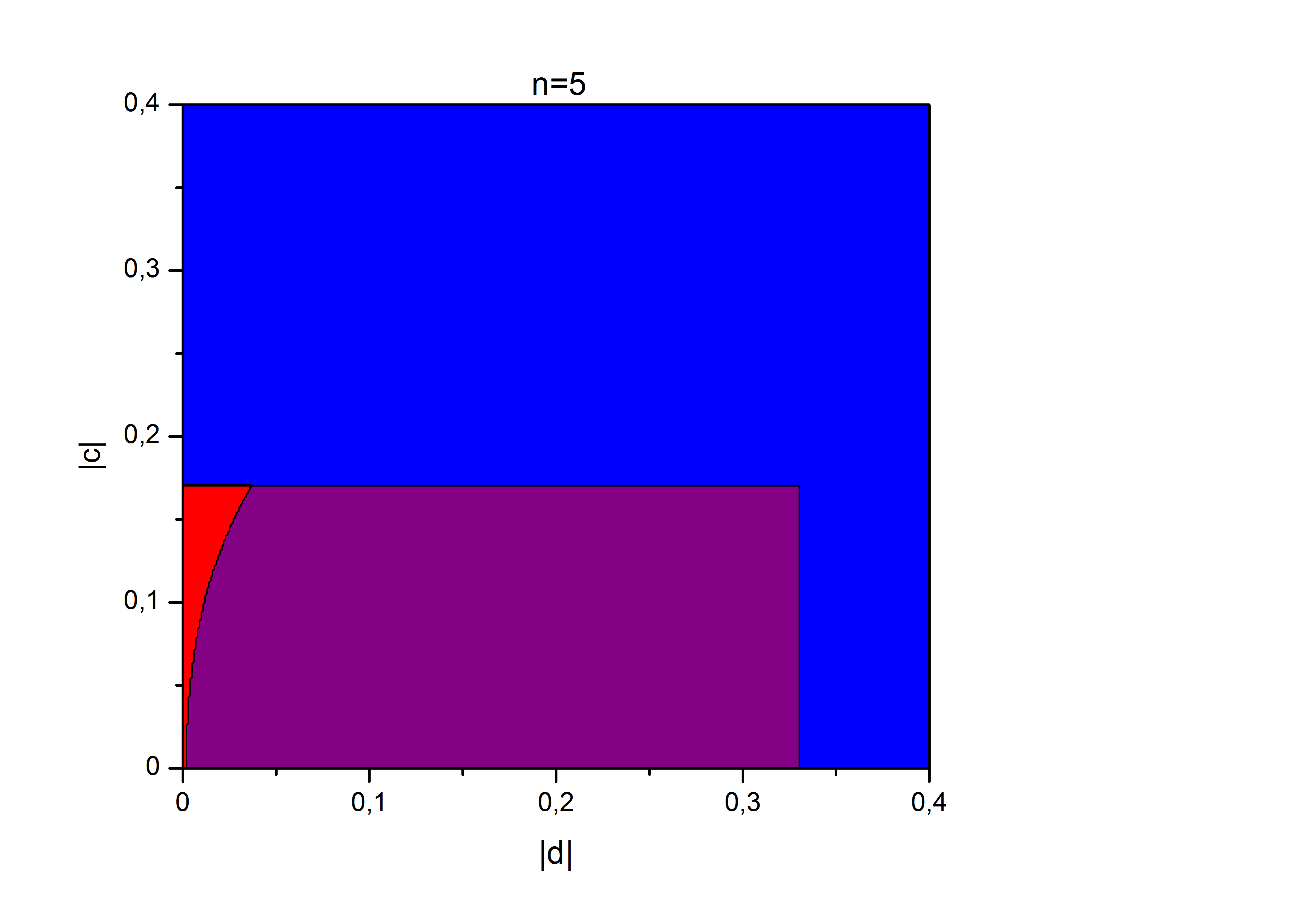}} b)\\
\end{minipage}
\vfill
\begin{minipage}[h]{0.47\linewidth}
\center{\includegraphics[width=1\linewidth]{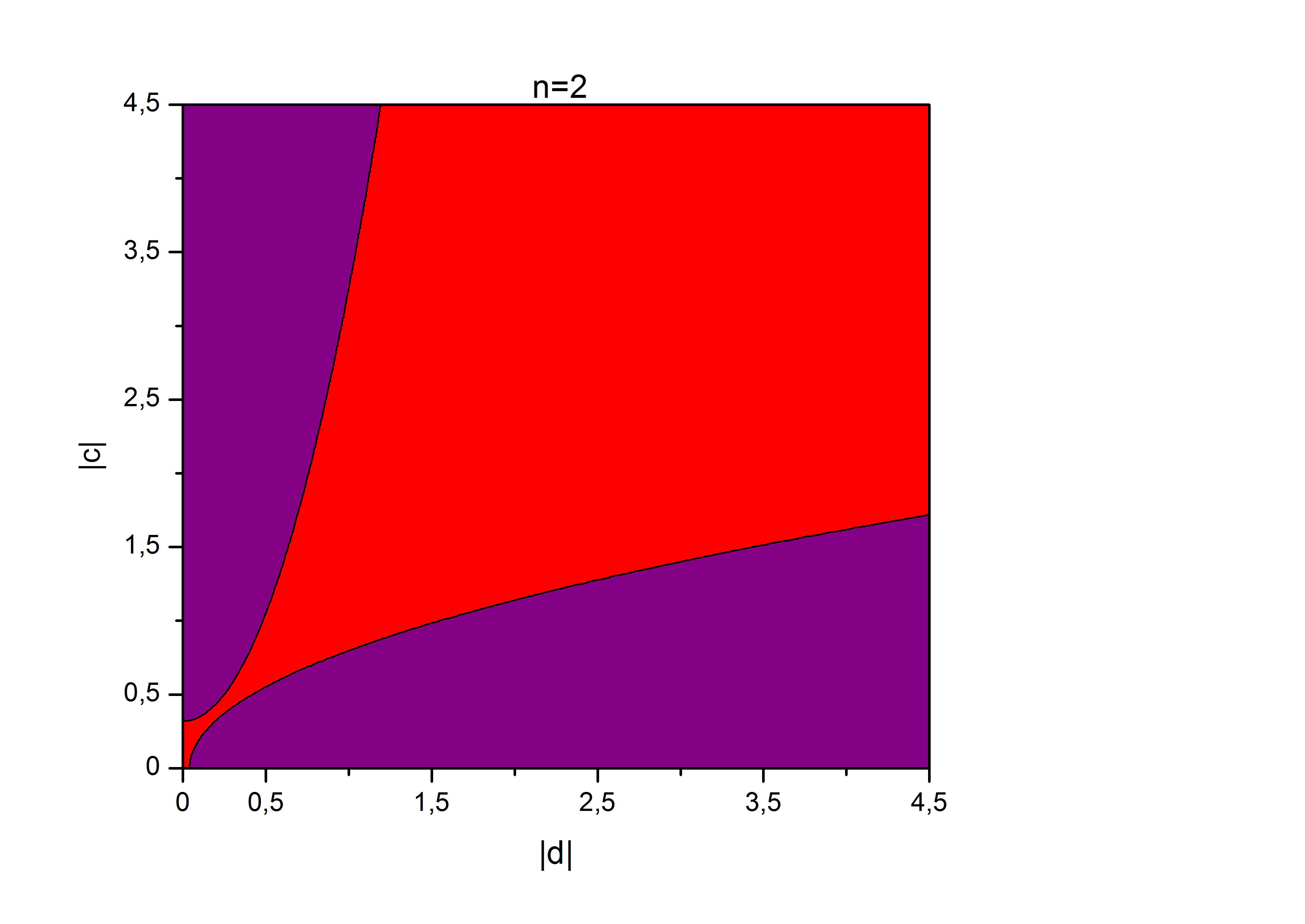}} c)\\
\end{minipage}
\hfill
\begin{minipage}[h]{0.47\linewidth}
\center{\includegraphics[width=1\linewidth]{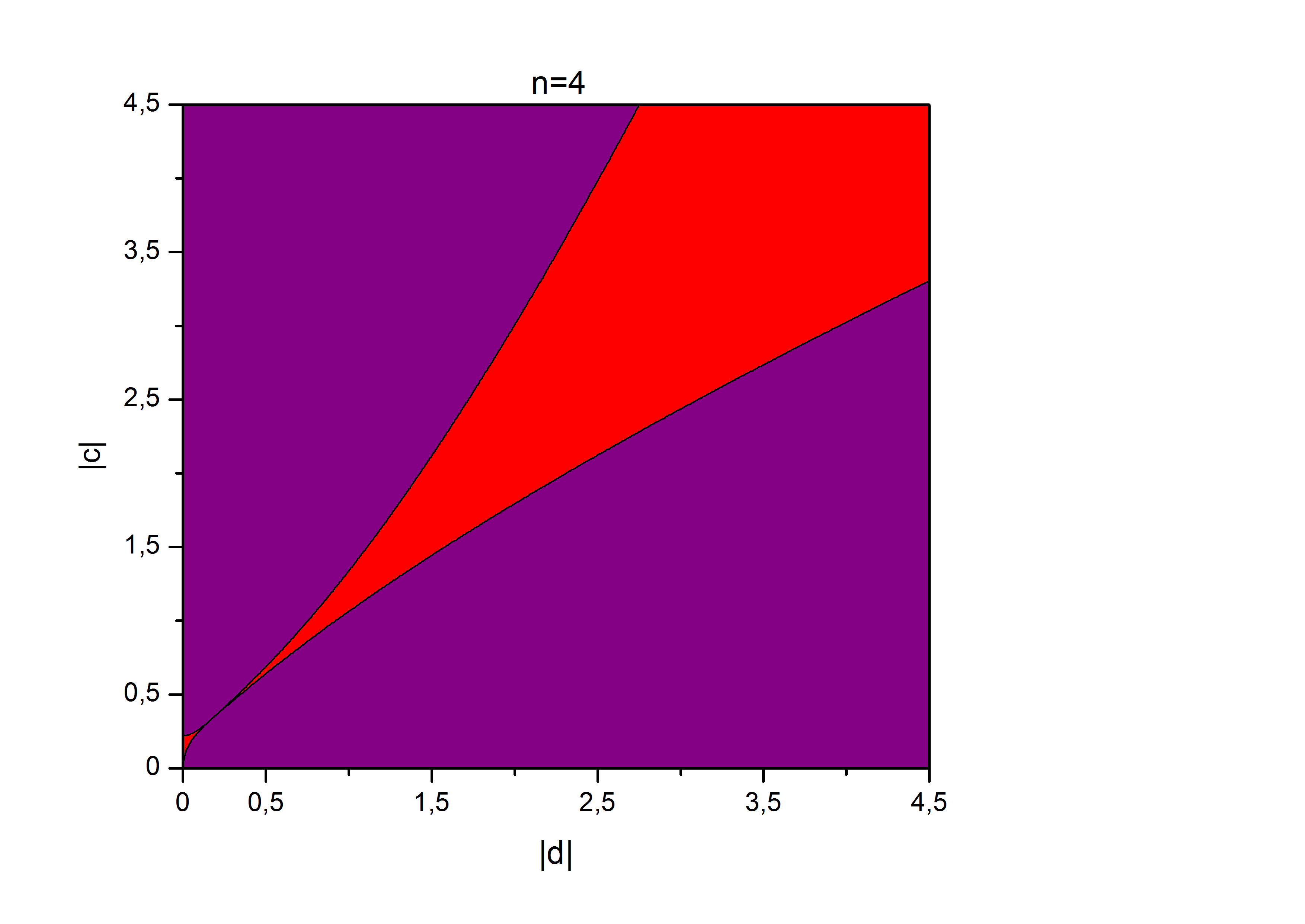}} d)\\
\end{minipage}
\caption{Domains of parameters $|c|$ and $|d|$ where density matrix $\rho_{X,n}$ describes separable (red color) and entangled (purple color). The blue domains correspond to set of parameters, where the matrix $\rho_{X,n}$ is not the density matrix. Parameters $a=0.33$ and $b=0.17$. Figures displays cases of different power $n$. a) $n=3$, b) $n=5$, c) $n=2$, d) $n=4$.}
\end{figure}

To find the form of density matrix $\rho_{X,n}$ produced by channel in case of arbitrary value of integer $n$ we make decomposition of matrix $\rho_X$
\be
\rho_X=SDS^{-1},
\ee
%(6)
where $D$ is diagonal matrix with eigenvalues of the matrix $\rho_X$ placed on the diagonal, and $S$ is unitary matrix containing the eigenvectors corresponding to these eigenvalues. The matrix $\rho_X$ has following eigenvalues and eigenvectors
\be
\begin{matrix}
\lambda_1=a+|d|\to \frac{1}{\sqrt{2}}\left(\begin{matrix}1 & 0 & 0 & \frac{d^*}{|d|}\end{matrix}\right)^T\\
\lambda_2=b+|c|\to \frac{1}{\sqrt{2}}\left(\begin{matrix}0 & 1 & \frac{c^*}{|c|} & 0\end{matrix}\right)^T\\
\lambda_3=b-|c|\to \frac{1}{\sqrt{2}}\left(\begin{matrix}0 & -\frac{c}{|c|} & 1 & 0\end{matrix}\right)^T\\
\lambda_4=a-|d|\to \frac{1}{\sqrt{2}}\left(\begin{matrix}-\frac{d}{|d|} & 0 & 0 & 1\end{matrix}\right)^T.
\end{matrix}
\ee
%(7)
Therefore the decomposition for the matrix $\rho_X$ can be written as
\be
\rho_{w,1}=SDS^{-1}=\frac{1}{\sqrt{2}}\begin{pmatrix}
1 & 0 & 0 & -\frac{d}{|d|}\\
0 & 1 & -\frac{c}{|c|} & 0\\
0 & \frac{c^*}{c} & 1 & 0\\
\frac{d^*}{|d|} & 0 & 0 & 1
\end{pmatrix}
\cdot
\begin{pmatrix}
\lambda_1 & 0 & 0 & 0\\
0 & \lambda_2 & 0 & 0\\
0 & 0 & \lambda_3 & 0\\
0 & 0 & 0 & \lambda_4
\end{pmatrix}
\cdot
\frac{1}{\sqrt{2}}\begin{pmatrix}
1 & 0 & 0 & \frac{d}{|d|}\\
0 & 1 & \frac{c}{|c|} & 0\\
0 & -\frac{c^*}{c} & 1 & 0\\
-\frac{d^*}{|d|} & 0 & 0 & 1
\end{pmatrix}.
\ee
%(8)
 This decomposition allows to find easily the result of raising matrix $\rho_X$ to power $n$
 \be
 \rho_X^n=(SDS^{-1})\cdot SDS^{-1}\cdot\ldots \cdot SDS^{-1}=SD^nS^{-1}
 \ee
To get the form of matrix $\rho_{X,n}$, one has to normalize the matrix $\rho_X^n$. Therefore the considered nonlinear channel gives the following matrix
 \begin{eqnarray}
 \rho_{X,n}=SD^nS^{-1}=[2(\lambda_1^n+\lambda_2^n+\lambda_3^n+\lambda_4^n)]^{-1}\cdot \nonumber \\ \cdot\begin{pmatrix}
 \lambda_1^n+\lambda_2^n & 0 & 0 & \left(\lambda_1^n-\lambda_2^n\right)\frac{d}{|d|}\\
 0 & \lambda_3^n+\lambda_4^n & \left(\lambda_3^n-\lambda_4^n\right)\frac{c}{|c|} & 0\\
 0 & \left(\lambda_3^n-\lambda_4^n\right)\frac{c^*}{|c|} & \lambda_3^n+\lambda_4^n & 0\\
 \left(\lambda_1^n-\lambda_2^n\right)\frac{d^*}{|d|} & 0 & 0 & \lambda_1^n+\lambda_2^n
 \end{pmatrix}.
 \end{eqnarray}
 %(11)
 From (11) it can be seen that the nonlinear channel gives the matrix of the form similar to the initial matrix $\rho_X$: the only non-zero elements are placed on the diagonal and anti-diagonal. To find the domains of parameters where the matrix $\rho_{X,n}$ describes states with different entanglement it is useful to introduce the notations
 \be
 A_n=\frac{\left(a+|d|\right)^n+\left(a-|d|\right)^n}{2(\lambda_1^n+\lambda_2^n+\lambda_3^n+\lambda_4^n)},
 \ee
 \be
 B_n=\frac{\left(b+|c|\right)^n+\left(b-|c|\right)^n}{2(\lambda_1^n+\lambda_2^n+\lambda_3^n+\lambda_4^n)},
 \ee
 \be
 C_n=\frac{\left(\left(b+|c|\right)^n-\left(b-|c|\right)^n\right)\frac{c}{|c|}}{2(\lambda_1^n+\lambda_2^n+\lambda_3^n+\lambda_4^n)},
 \ee
 \be
 D_n=\frac{\left[\left(a+|d|\right)^n-\left(a-|d|\right)^n\right]\frac{d}{|d|}}{2(\lambda_1^n+\lambda_2^n+\lambda_3^n+\lambda_4^n)}.
 \ee
 %(12)
Thus this task becomes equivalent to the question about entanglement of state corresponding to the density matrix $\rho_X$. The conditions of positivity of matrix $\rho_{X,n}$ are $A_n\ge|D_n|$ and $B_n\ge|C_n|$. The two cases of odd and even integers $n$ have to be separated. If integer $n$ is even, the matrix $\rho_{X,n}$ is the density matrix for arbitrary set of parameters $a$, $b$, $c$, $d$. In case of odd integer $n$ there is domain of parameters where the matrix $\rho_{X,n}$ is negative. The condition of positivity for odd integers $n$ can be rewritten in the following form
$$
\left[
\begin{array}{rcl}
a\ge|d|, b\ge|c|, ((a+|d|)^n+(a-|d|)^n+(b+|c|)^n+(b-|c|)^n)>0,\\
a\le|d|, b\le|c|, ((a+|d|)^n+(a-|d|)^n+(b+|c|)^n+(b-|c|)^n)<0.\\
\end{array}
\right.
$$
%(13)

According to Peres-Horodecki criterion the new state is entangled when at least one of the conditions $A_n\ge |C_n|$ or $B_n\ge |D_n|$ is violated. Figure 2 displays the domains of absolute values of $c$ and $d$ where matrix $\rho_{X,n}$ describes the state with different properties of entanglement.

If the initial state is Werner state with parameter $p$ the conditions of positivity of matrix $\rho_{X,n}$ are turned to conditions on parameter $p$. In case of odd $n$ parameter $p \in \left[1/3;1\right]$. If integer $n$ is even the matrix $\rho_{X,n}$ is positive for all real values of parameter $p$. The entangled state is described by matrix $\rho_{X,n}$ in domain $(1-\frac{4}{\sqrt[n]{3}+3}; 1]$ of parameter $p$ for odd integer $n$ and in domain $(p<1+\frac{4}{\sqrt[n]{3}-3})\cup(p>1-\frac{4}{\sqrt[n]{3}+3})$ if integer $n$ is even.

\section{Quantum tomography for nonlinear channels}
\pst
Quantum tomography for spin state shows the probability distribution of spin projections on selected directions. If the direction $\overrightarrow{n}$ corresponds to unitary matrix
\be
u=\begin{pmatrix}
\cos{\left(\frac{\theta}{2}\right)}e^{\frac{i}{2}(\varphi+\psi)}&\sin{\left(\frac{\theta}{2}\right)}e^{\frac{i}{2}(\varphi-\psi)}\\
-\sin{\left(\frac{\theta}{2}\right)}e^{-\frac{i}{2}(\varphi-\psi)}&\cos{\left(\frac{\theta}{2}\right)e^{-\frac{i}{2}(\varphi+\psi)}}
\end{pmatrix},
\ee
depending on Euler's angles $\theta$, $\varphi$, $\psi$, the quantum tomogram for spin state with density matrix $\hat{\rho}$ is produced by formula
\be
W(m,\overrightarrow{n})=\langle{m}|u\rho u^+|{m}\rangle,
\ee
where $m$ is spin projection on the direction $\overrightarrow{n}$. The quantum tomogram for two qubit system and directions $\overrightarrow{a}$, $\overrightarrow{b}$ is given by formula
\be
W(m_1,\overrightarrow{a},m_2,\overrightarrow{b})=\langle{j_1,m_1,j_2,m_2}|{U\rho U^+}|{j_1,m_1,j_2,m_2}\rangle,
\ee
where $U=u_1\otimes{u_2}$, the product being tensor product of unitary matrices $u_1$, $u_2$ corresponding to directions $\overrightarrow{a}$, $\overrightarrow{b}$, and $m_1$, $m_2$ are spin projections on these directions.

The general form of matrices, produced by considered nonlinear channel, is X-matrix with specific matrix elements for each power $n$ of channel. Thus, it is possible to write formula for quantum tomogram in following form
 \begin{eqnarray}
        W(\uparrow,\overrightarrow{a},\uparrow,\overrightarrow{b})=W(\downarrow,\overrightarrow{a},\downarrow,\overrightarrow{b})=A_n\cdot f_{+}\left(\theta_a,\theta_b\right)+\nonumber\\+B_n\cdot f_{-}\left(\theta_a,\theta_b\right)+\psi\left(C_n,D_n,\theta_a,\theta_b,\Psi_a,\Psi_b\right),
        \nonumber \\
        W(\uparrow,\overrightarrow{a},\downarrow,\overrightarrow{b})=W(\downarrow,\overrightarrow{a},\uparrow,\overrightarrow{b})=A_n\cdot f_{-}\left(\theta_a,\theta_b\right)+\nonumber\\+B_n\cdot f_{+}\left(\theta_a,\theta_b\right)-\psi\left(C_n,D_n,\theta_a,\theta_b,\Psi_a,\Psi_b\right),
\end{eqnarray}
        where $A_n$, $B_n$, $C_n$, $D_n$ given by (12)-(15), $f_{+}\left(\theta_a,\theta_b\right)=\cos^2{\left(\frac{\theta_a}{2}\right)}\cos^2{\left(\frac{\theta_b}{2}\right)}+\sin^2{\left(\frac{\theta_a}{2}\right)}\sin^2{\left(\frac{\theta_b}{2}\right)}$, $f_{-}\left(\theta_a,\theta_b\right)=\cos^2{\left(\frac{\theta_a}{2}\right)}\sin^2{\left(\frac{\theta_b}{2}\right)}+\sin^2{\left(\frac{\theta_a}{2}\right)}\cos^2{\left(\frac{\theta_b}{2}\right)}$ and $\psi\left(C_n,D_n,\theta_a,\theta_b,\Psi_a,\Psi_b\right)=\frac{1}{2}\sin{\theta_a}\sin{\theta_b}\cdot Re\left(ce^{i(\Psi_a-\Psi_b)}+de^{i(\Psi_a+\Psi_b)}\right)$. The formulas are written for arbitrary integer $n$ determining the nonlinear channel. The only difference between cases of different $n$ is due to the matrix elements $A_n$, $B_n$, $C_n$, $D_n$.

In particular, quantum tomogram for directions $\overrightarrow{n_1}$ and $\overrightarrow{n_2}$ in case of initial Werner state takes form
\begin{eqnarray}
W(\uparrow,\overrightarrow{n_1},\uparrow,\overrightarrow{n_2})=W(\downarrow,\overrightarrow{n_1},\downarrow,\overrightarrow{n_2})=A(p)\left(\cos^2{\left(\frac{\theta_1}{2}\right)}\cos^2{\left(\frac{\theta_2}{2}\right)}+\sin^2{\left(\frac{\theta_1}{2}\right)}\sin^2{\left(\frac{\theta_2}{2}\right)}\right)+\nonumber\\+B(p)\left(\sin^2{\left(\frac{\theta_1}{2}\right)}\cos^2{\left(\frac{\theta_2}{2}\right)}+\cos^2{\left(\frac{\theta_1}{2}\right)}\sin^2{\left(\frac{\theta_2}{2}\right)}\right)+\frac{C(p)}{2}\sin\left({\theta_1}\right)\sin\left({\theta_2}\right)\cos\left(\psi_1+\psi_2\right),
\end{eqnarray}
%(15)
\begin{eqnarray}
W(\uparrow,\overrightarrow{n_1},\downarrow,\overrightarrow{n_2})=W(\downarrow,\overrightarrow{n_1},\uparrow,\overrightarrow{n_2})=A(p)\left(\sin^2{\left(\frac{\theta_1}{2}\right)}\cos^2{\left(\frac{\theta_2}{2}\right)}+\cos^2{\left(\frac{\theta_1}{2}\right)}\sin^2{\left(\frac{\theta_2}{2}\right)}\right)+\nonumber\\+B(p)\left(\cos^2{\left(\frac{\theta_1}{2}\right)}\cos^2{\left(\frac{\theta_2}{2}\right)}+\sin^2{\left(\frac{\theta_1}{2}\right)}\sin^2{\left(\frac{\theta_2}{2}\right)}\right)-\frac{C(p)}{2}\sin\left({\theta_1}\right)\sin\left({\theta_2}\right)\cos\left(\psi_1+\psi_2\right),
\end{eqnarray}
%(16)
where $A(p)=\frac{1}{2}\frac{(1-p)^n+(1+3p)^n}{3(1-p)^n+(1+3p)^n}$, $B(p)=\frac{(1-p)^n}{3(1-p)^n+(1+3p)^n}$ and $C(p)=A(p)-B(p)$.

\section{Entropic inequalities for nonlinear channels $\rho_X\to\Phi(\rho_X)$}
\pst
The von Neumann entropy of arbitrary system with density matrix $\rho$ is defined as follows~\cite{7}
\begin{equation}
S=-\mathrm{Tr}\left(\rho \ln{\rho}\right).
\end{equation}
Let us consider the case of two qubit system with qubits 1, 2. In case of two qubit system state $\rho(1,2)$ there are entropy $S(1,2)$ for system and entropies $S(1),S(2)$ for subsystems corresponding to density matrices $\rho(1)=\mathrm{Tr_2}\rho(1,2)$ and $\rho(2)=\mathrm{Tr_1}\rho(1,2)$
\begin{eqnarray}
S(1,2)=-\mathrm{Tr}\left[\rho(1,2)\ln{\rho(1,2)}\right]\nonumber \\S(k)=-\mathrm{Tr}\left[\rho(k)\ln{\rho(k)}\right], k=(1,2)
\end{eqnarray}
For X-state of two qubits we have $\rho(1,2)=\rho_X$. In case of matrices produced by nonlinear channel from $\rho_X$
\begin{equation}
\begin{matrix}
\rho(1)=\left(
\begin{matrix}
\frac{1}{2} & 0\\
0 & \frac{1}{2} 
\end{matrix}\right) && \rho(2)=\left(
\begin{matrix}
\frac{1}{2} & 0\\
0 & \frac{1}{2} 
\end{matrix}\right)   	
\end{matrix}
\end{equation}
and $S(1)=S(2)=\ln{2}$. Entropy of system is $S(1,2)=-\sum_{k=1}^{4}\lambda_k\ln{\lambda_k}$, where $\lambda_k$ are $A_n\pm |D_n|$, $B_n\pm |C_n|$ - eigenvalues of $\rho_{X,n}$.
The quantum mutual information of system with subsystems $1,2$ is defined as
\begin{equation}
I_N=S(1)+S(2)-S(1,2).
\end{equation}
In case of initial Werner state the information $I_N(p)$ is
\begin{eqnarray}
I_N(p)=\ln{4}+\ln{[(1+3p)^n+3(1-p)^n]}-\nonumber\\-\frac{(1+3p)^n\ln{(1+3p)^n}+3(1-p)^n\ln{(1-p)^n}}{(1+3p)^n+3(1-p)^n}.
\end{eqnarray}
It has maximum at $p=1$ where $I_N=\ln{4}$. The quantum mutual information $I_N(p)$ increases after action of nonlinear channel on initaial state. With growing integer $n$ determining the nonlinear channel the information is increasing.

\begin{figure}[ht]
\begin{minipage}[h]{0.47\linewidth}
\center{\includegraphics[width=1\linewidth]{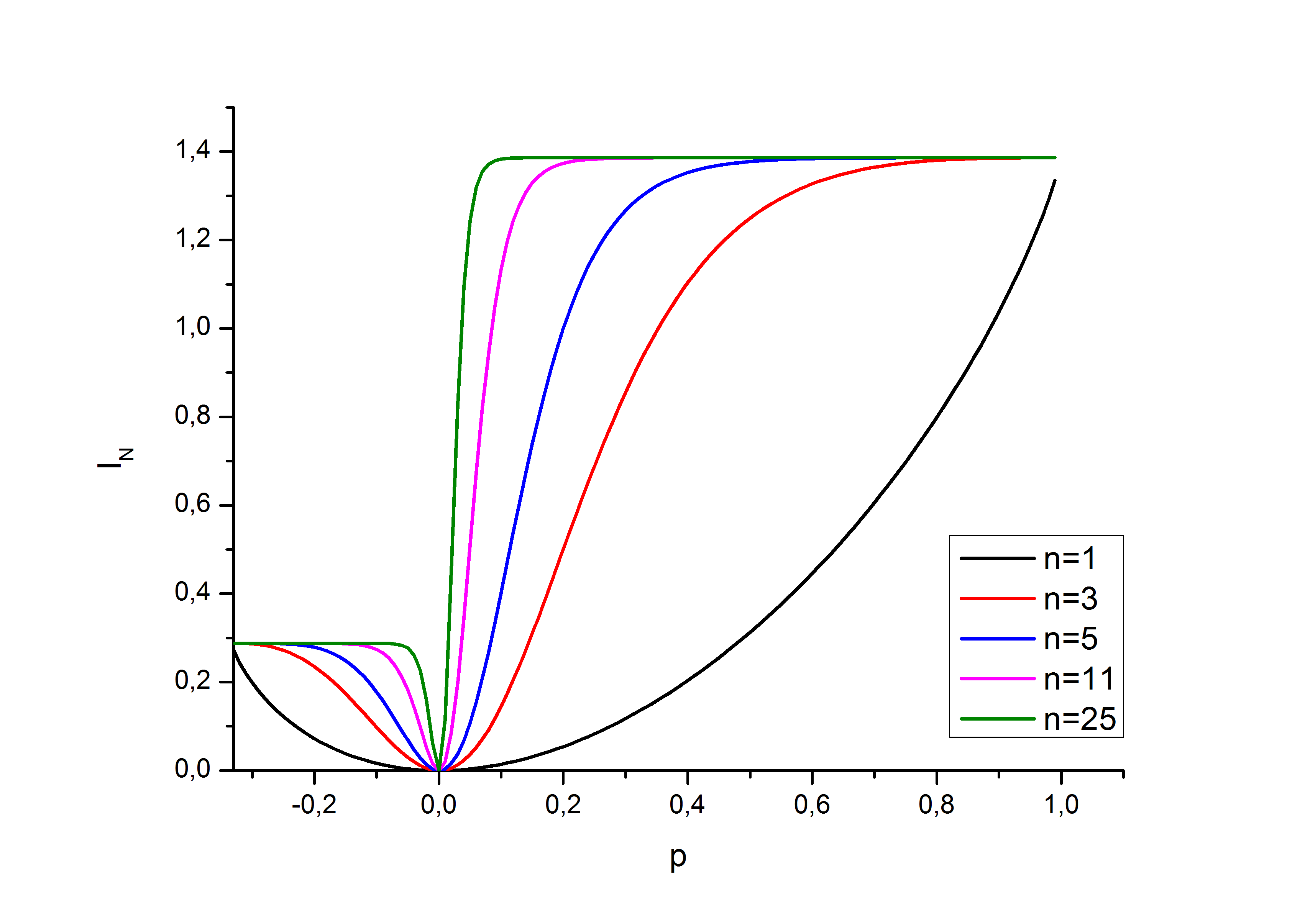}} a)\\
\end{minipage}
\hfill
\begin{minipage}[h]{0.47\linewidth}
\center{\includegraphics[width=1\linewidth]{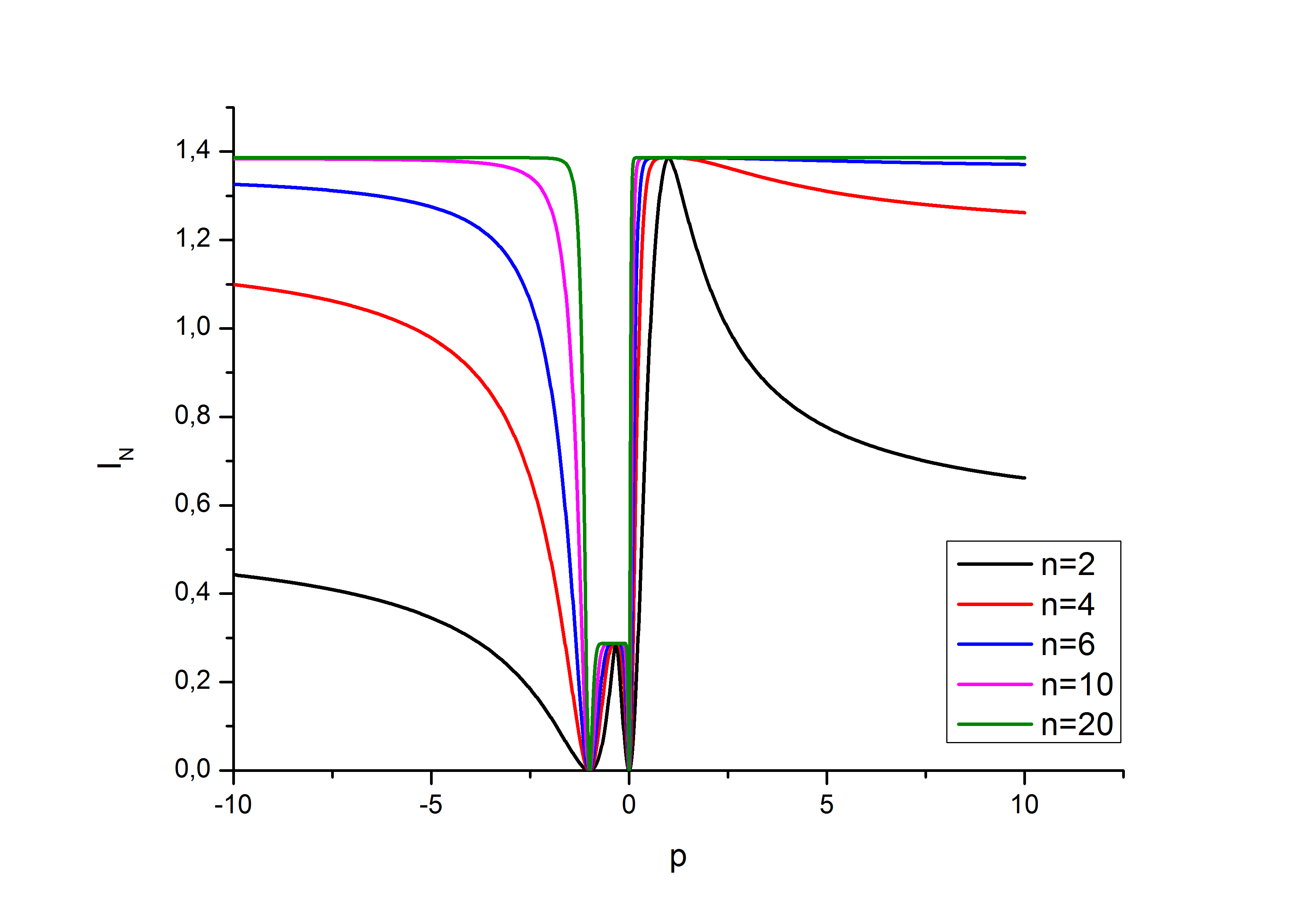}} b)\\
\end{minipage}
\caption{Mutual information for different $n$. Figures a) and b) correspond to cases of odd and even $n$, respectively.}
\end{figure}

Tomographic Shannon entropy~\cite{24} is associated with tomographic probability distribution. For subsystems of considered two qubit systems the marginal distributions are
\begin{eqnarray}
	W_1(\uparrow,\overrightarrow{a})=W(\uparrow,\overrightarrow{a},\uparrow,\overrightarrow{b})+W(\uparrow,\overrightarrow{a},\downarrow,\overrightarrow{b}),\nonumber\\W_1(\downarrow,\overrightarrow{a})=W(\downarrow,\overrightarrow{a},\uparrow,\overrightarrow{b})+W(\downarrow,\overrightarrow{a},\downarrow,\overrightarrow{b}),\nonumber\\\nonumber\\W_2(\uparrow,\overrightarrow{b})=W(\uparrow,\overrightarrow{a},\uparrow,\overrightarrow{b})+W(\downarrow,\overrightarrow{a},\uparrow,\overrightarrow{b}),\nonumber\\W_2(\downarrow,\overrightarrow{b})=W(\uparrow,\overrightarrow{a},\downarrow,\overrightarrow{b})+W(\downarrow,\overrightarrow{a},\downarrow,\overrightarrow{b}).
\end{eqnarray}
Tomographic entropies of qubit subsystems are
\begin{eqnarray}
	H(1)=-W_1(\uparrow,\overrightarrow{a})\ln{W_1(\uparrow,\overrightarrow{a})}-W_1(\downarrow,\overrightarrow{a})\ln{W_1(\downarrow,\overrightarrow{a})},\nonumber\\H(2)=-W_2(\uparrow,\overrightarrow{b})\ln{W_2(\uparrow,\overrightarrow{b})}-W_2(\downarrow,\overrightarrow{b})\ln{W_2(\downarrow,\overrightarrow{b})}.
\end{eqnarray}
Tomographic entropy for two qubit system state is given by formula
\begin{eqnarray}
H(1,2)=-W(\uparrow,\overrightarrow{a},\uparrow,\overrightarrow{b})\ln{W(\uparrow,\overrightarrow{a},\uparrow,\overrightarrow{b})}-\nonumber\\-W(\downarrow,\overrightarrow{a},\uparrow,\overrightarrow{b})\ln{W(\downarrow,\overrightarrow{a},\uparrow,\overrightarrow{b})}-\nonumber\\-W(\uparrow,\overrightarrow{a},\downarrow,\overrightarrow{b})\ln{W(\uparrow,\overrightarrow{a},\downarrow,\overrightarrow{b})}-\nonumber\\-W(\downarrow,\overrightarrow{a},\downarrow,\overrightarrow{b})\ln{W(\downarrow,\overrightarrow{a},\downarrow,\overrightarrow{b})}.
\end{eqnarray}
For states $\rho_{X,n}$ entropies of subsystems equal to $H(1)=H(2)=\ln{2}$.

\begin{figure}[ht]
\begin{minipage}[h]{0.47\linewidth}
\center{\includegraphics[width=1\linewidth]{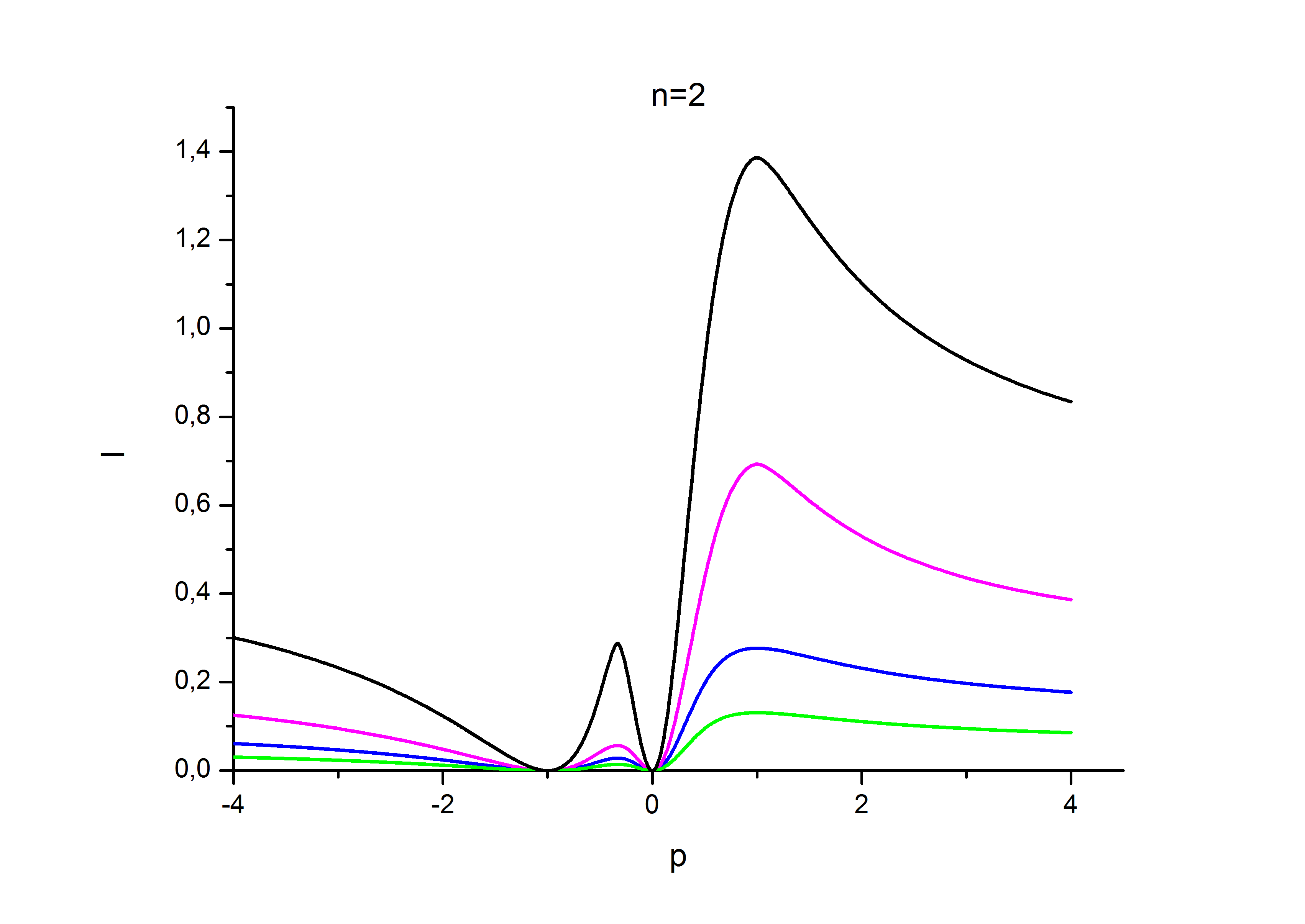}} a)\\
\end{minipage}
\hfill
\begin{minipage}[h]{0.47\linewidth}
\center{\includegraphics[width=1\linewidth]{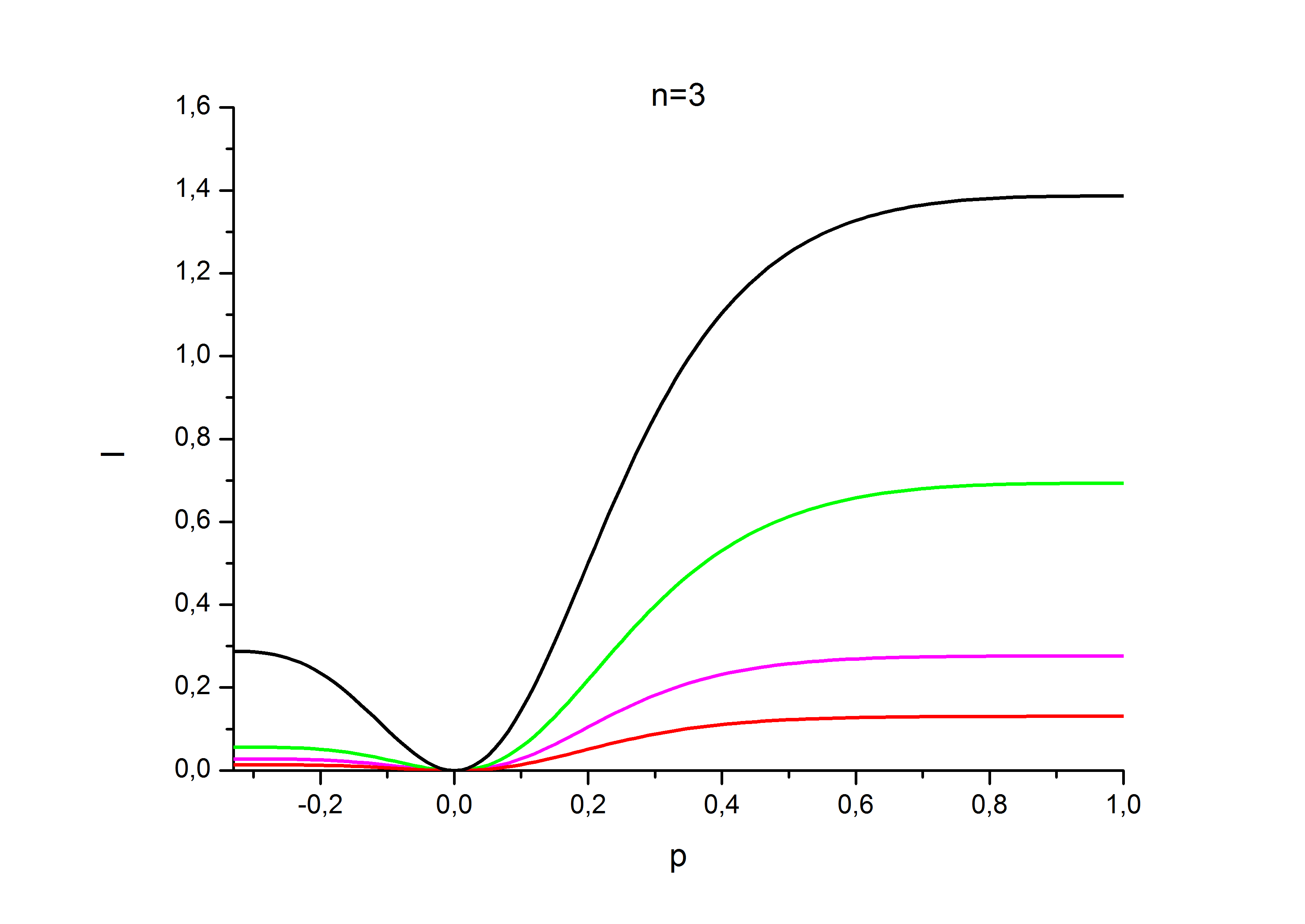}} b)\\
\end{minipage}
\caption{Mutual information (black line) and Shannon information for few sets of directions (coloured lines). Figures a) and b) correspond to cases of $n=2$ and $n=3$, respectively.}
\end{figure}

Shannon information for two qubit system is nonnegative and it is given by formula
\begin{equation}
I_S=H(1)+H(2)-H(1,2).
\end{equation}
The relation between two definitions of information (14) and (19) is 
\begin{equation}
I_S\le I_N
\end{equation}

For both mutual and Shannon informations the nonnegativity conditions exist $I_N\ge 0$, $I_S\ge 0$.

\section{Negativity and concurrence}
\pst

We use such characteristics as negativity and concurrence~\cite{16,17} to analyze action of considered quantum channel on entanglement of X-state. Negativity for system having density matrix $\rho$ is defined as follows
\begin{equation}
N=\mathrm{Tr}|\rho^{ppt}|,
\end{equation}
where $\rho^{ppt}$ - ppt-transformed matrix $\rho$. Partial transpose density matrix for separable state remains nonnegative and saves the property of $\mathrm{Tr}\rho=1$. Negativity $N$ of entangled state deviates from this value. Negativity gets greater for more entangled state.

\begin{figure}[ht]
\begin{minipage}[h]{0.47\linewidth}
\center{\includegraphics[width=1\linewidth]{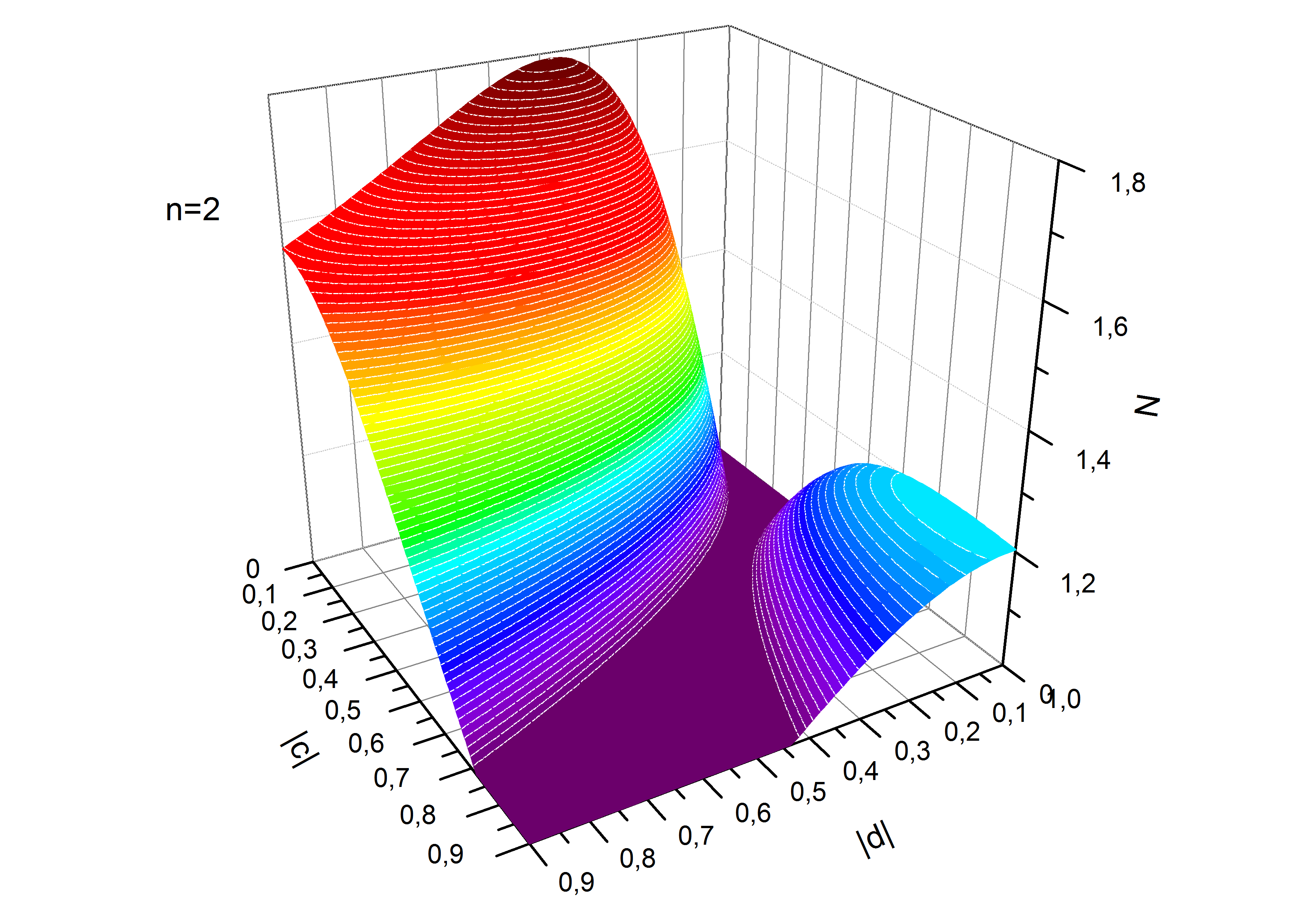}} a)\\
\end{minipage}
\hfill
\begin{minipage}[h]{0.47\linewidth}
\center{\includegraphics[width=1\linewidth]{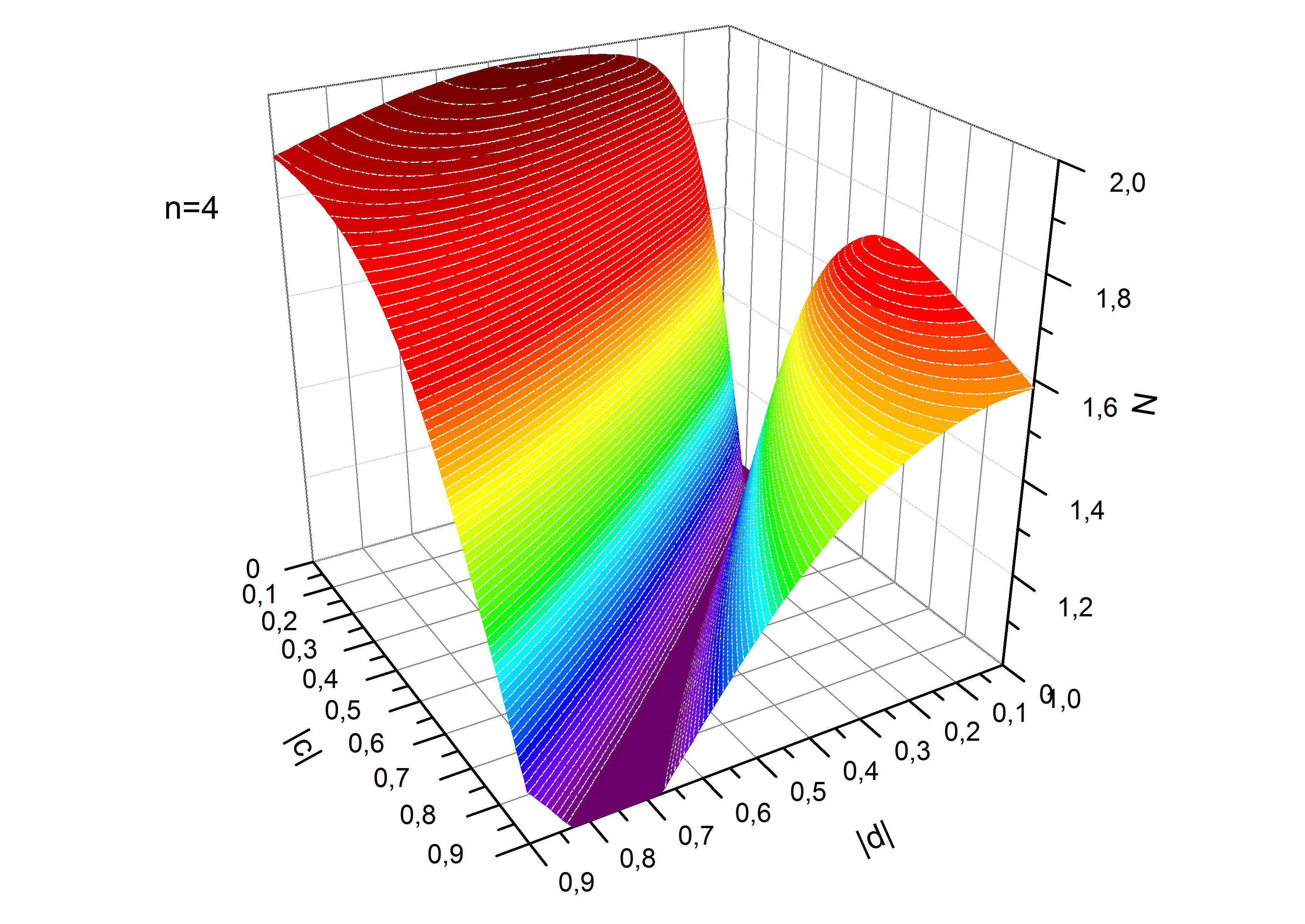}} b)\\
\end{minipage}
\vfill
\begin{minipage}[h]{0.47\linewidth}
\center{\includegraphics[width=1\linewidth]{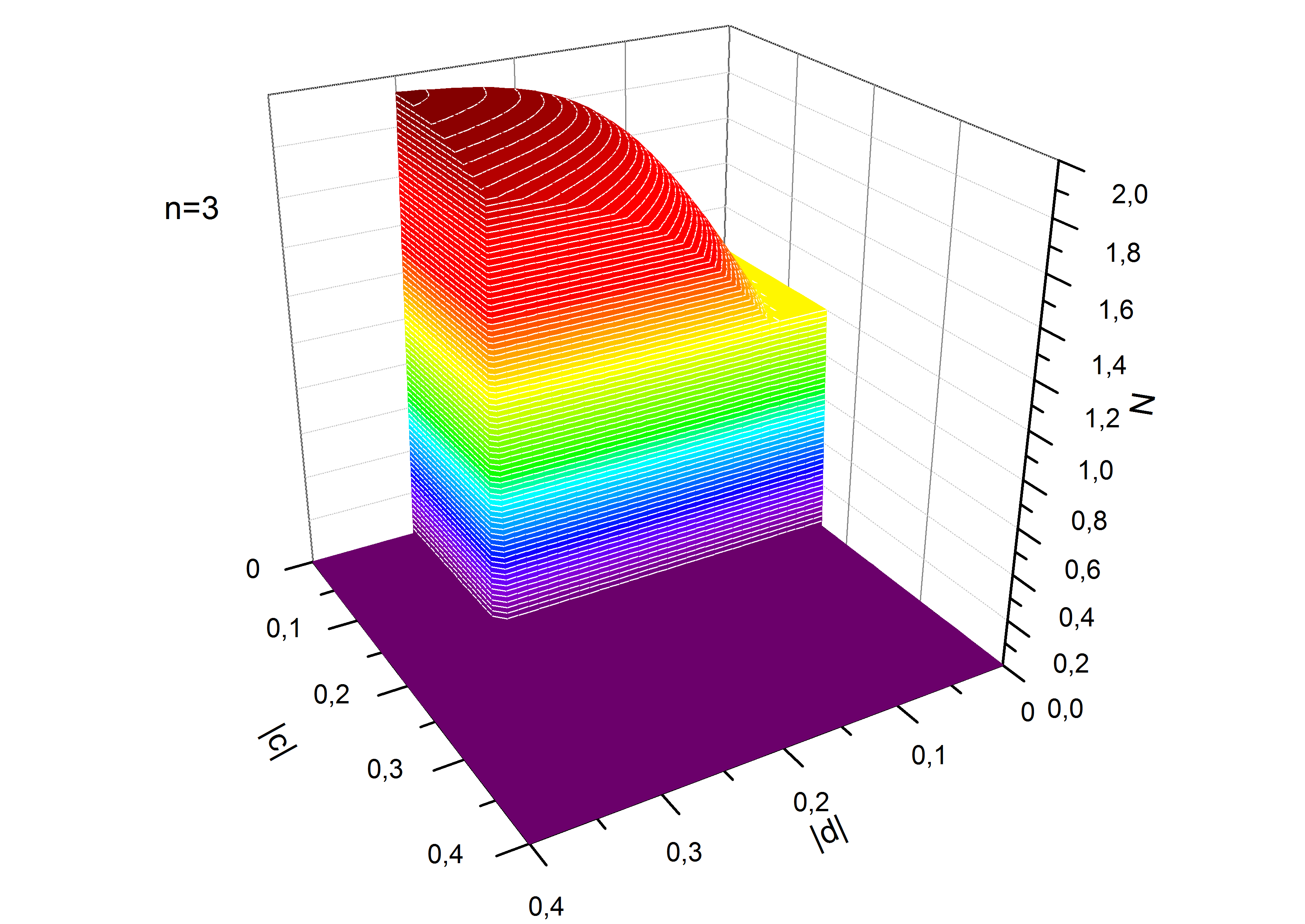}} c)\\
\end{minipage}
\hfill
\begin{minipage}[h]{0.47\linewidth}
\center{\includegraphics[width=1\linewidth]{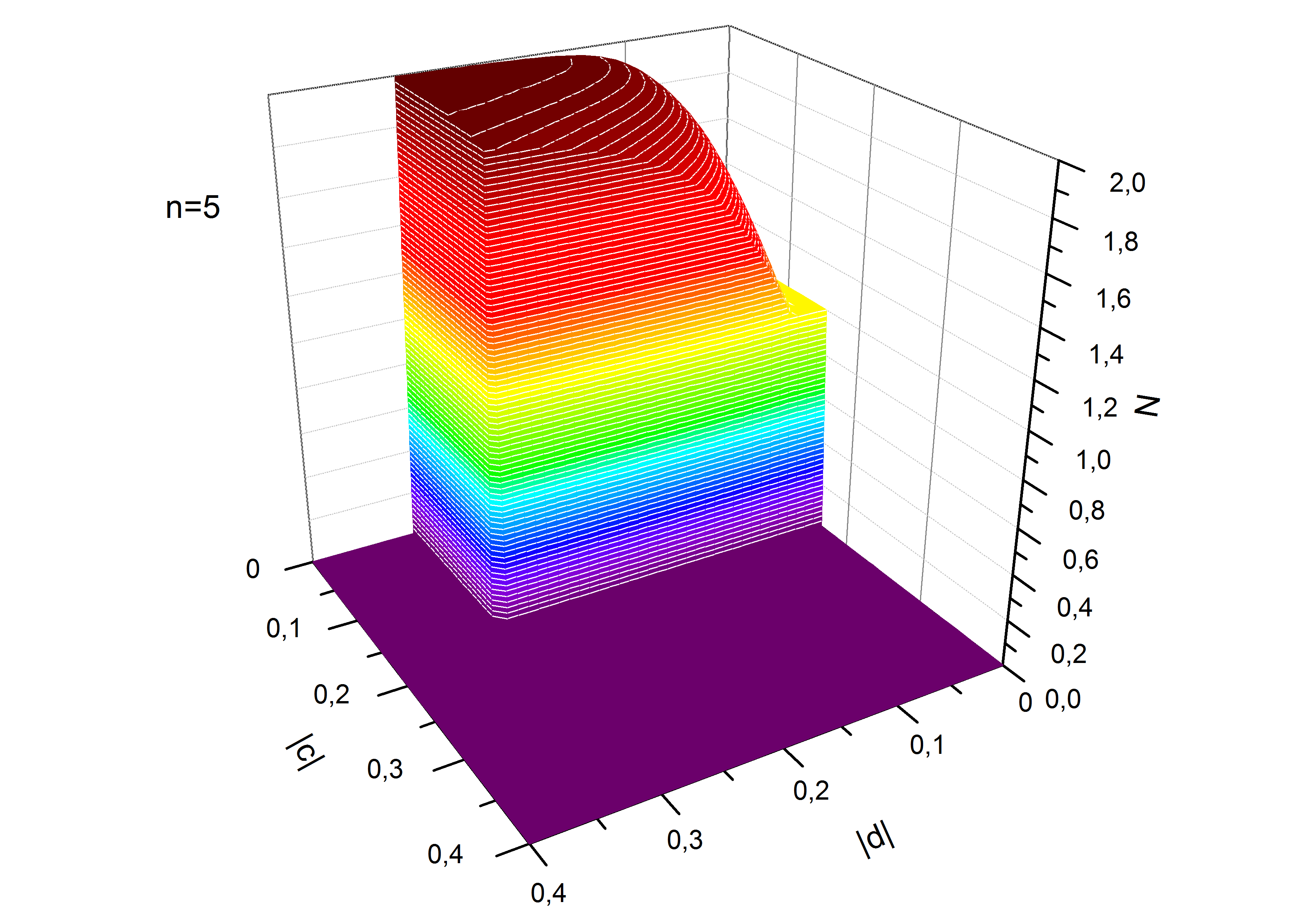}} d)\\
\end{minipage}
\caption{Negativity for states poduced by quantum channel with a) $n=2$, b) $n=4$, c) $n=3$ and d) $n=5$ for different values of $|c|$ and $|d|$. Parameters $a=0.33$ and $b=0.17$. Level $N=0$ correspond the case when matrix $\rho_{X,n}$ is negative. Level $N=1$ is for separable state. If the state is entangled then $N>1$.}
\end{figure}

Figure 5 displays the negativity for density matrices produced by channel $\rho_X\to\rho_{X,n}$ for different cases of power $n$. The surfaces show dependences of negativity on the absolute values of parameters $c$ and $d$ of matrix $\rho_X$. For odd $n$ behaviors of surfaces distinguish significantly from cases of even $n$. However, one can see that for both cases the domains of parameters $c$ and $d$ ,where new state is entangled, are expanding with power $n$. The value of negativity is growing with integer $n$.
Concurrence for two qubit state with density matrix $\rho$ is defined by formula
\begin{equation}
C=max(0,\sqrt{\lambda_1}-\sqrt{\lambda_2}-\sqrt{\lambda_3}-\sqrt{\lambda_4}),
\end{equation}
where $\lambda_k, k=1,2,3,4$ - are eigenvalues of matrix $R=\rho\rho_C$, and $\lambda_1$ has the maximum value. Matrix $\rho_C$ is result of action of spin flip operation on matrix $\rho$
\begin{equation}
\rho_C=(\sigma_y\otimes\sigma_y)\rho^*(\sigma_y\otimes\sigma_y),
\end{equation}
where $\sigma_y$ is Pauli matrix.
\begin{figure}[ht]
\begin{minipage}[h]{0.47\linewidth}
\center{\includegraphics[width=1\linewidth]{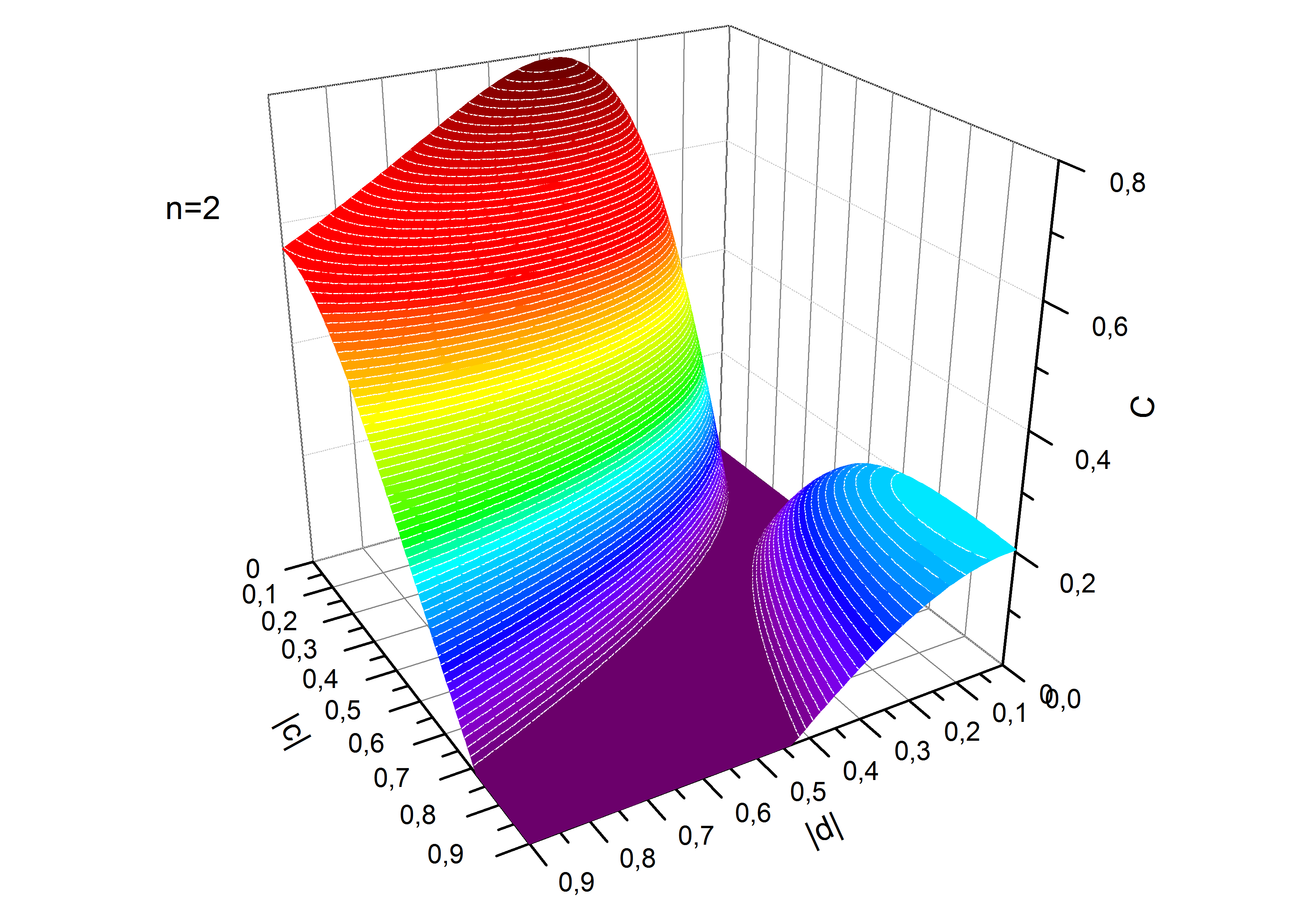}} a)\\
\end{minipage}
\hfill
\begin{minipage}[h]{0.47\linewidth}
\center{\includegraphics[width=1\linewidth]{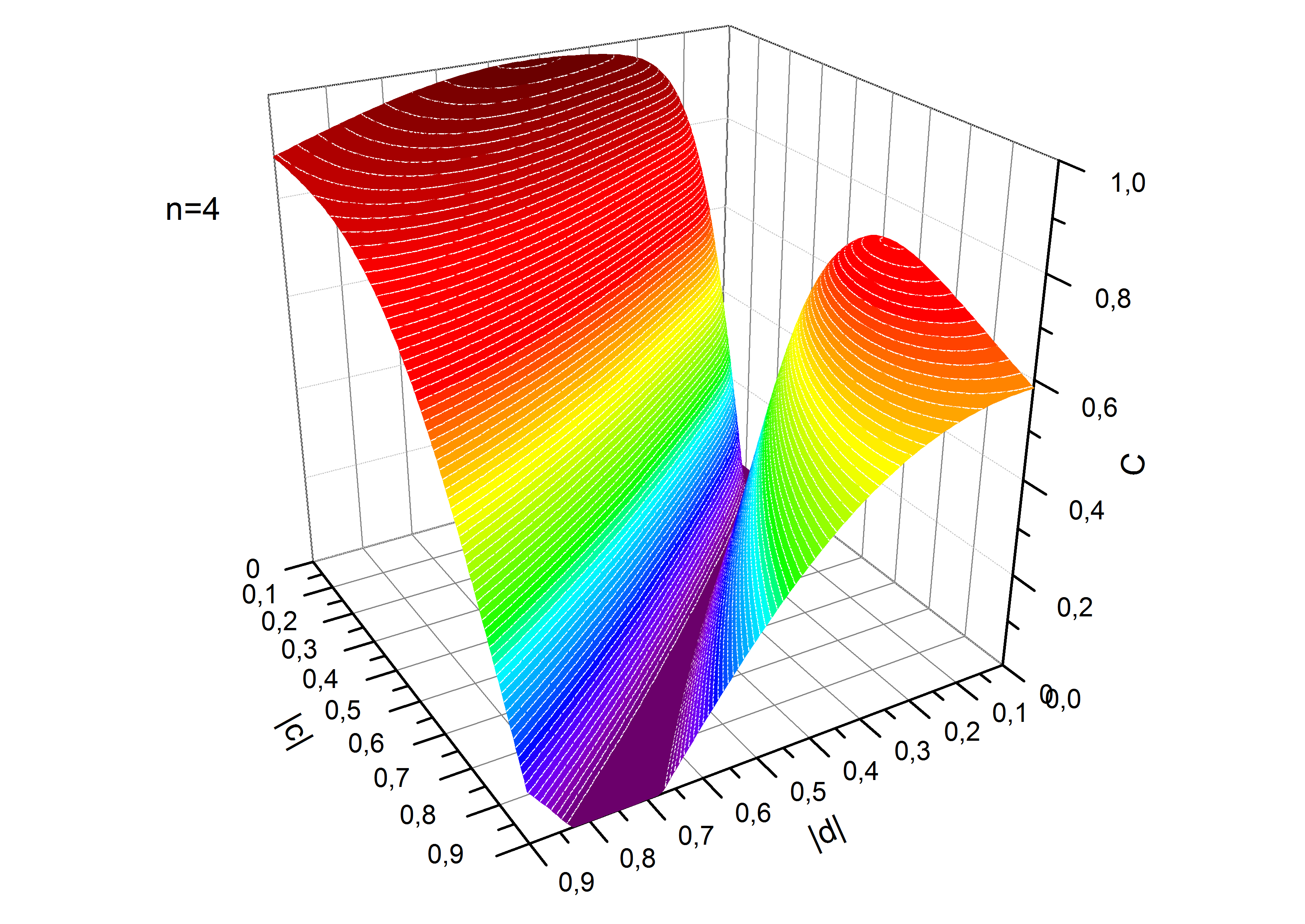}} b)\\
\end{minipage}
\vfill
\begin{minipage}[h]{0.47\linewidth}
\center{\includegraphics[width=1\linewidth]{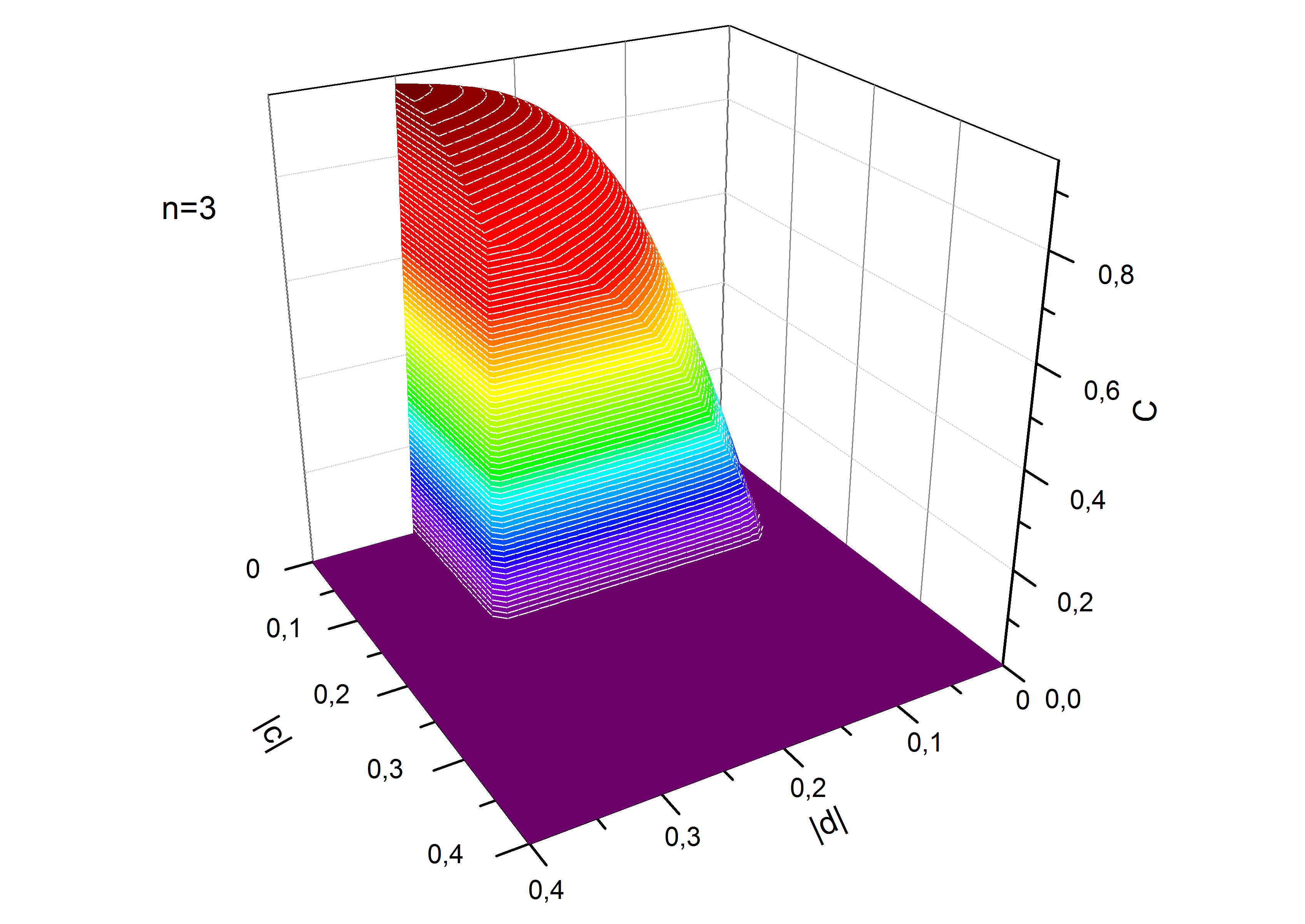}} c)\\
\end{minipage}
\hfill
\begin{minipage}[h]{0.47\linewidth}
\center{\includegraphics[width=1\linewidth]{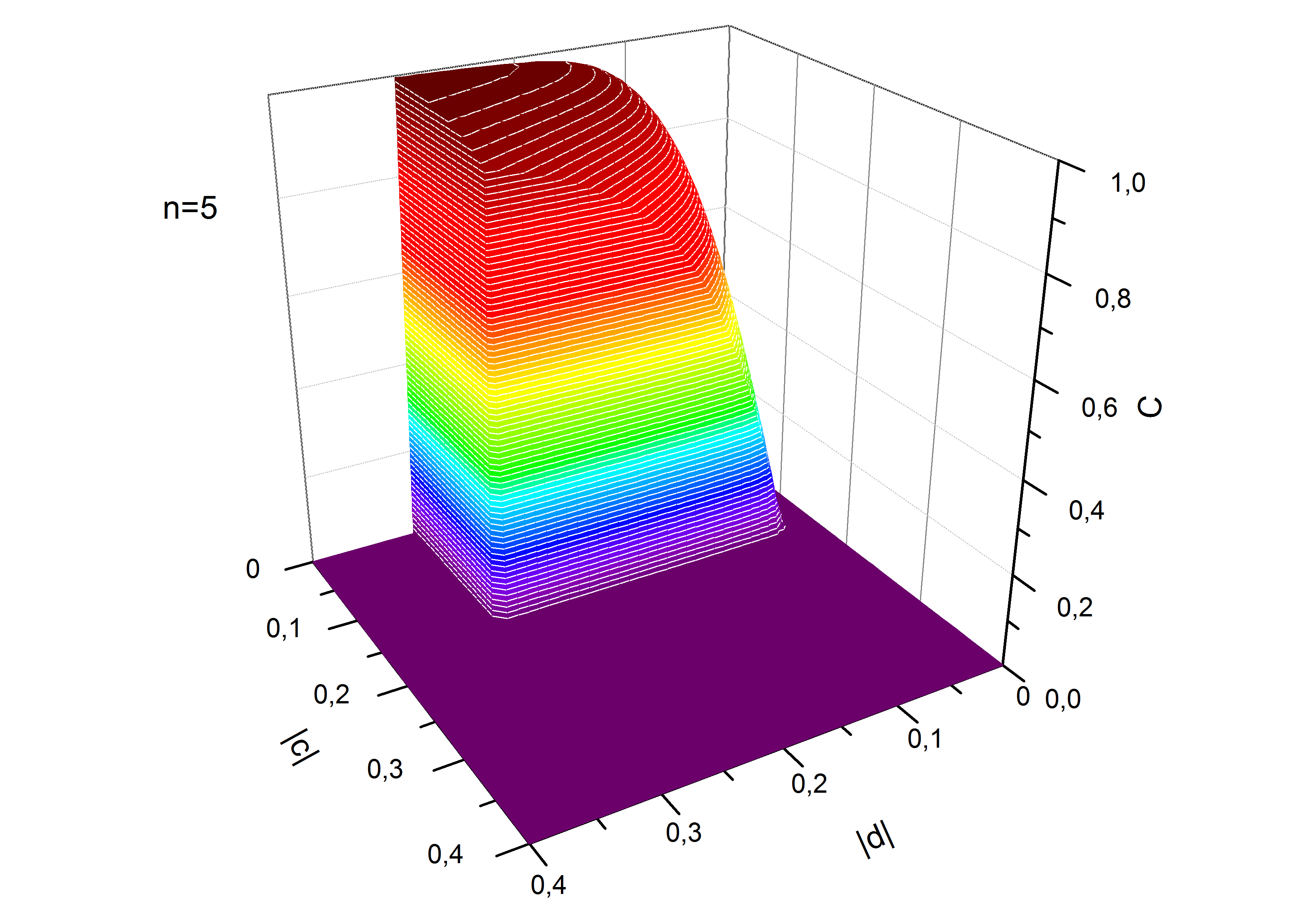}} d)\\
\end{minipage}
\caption{Concurrence for states poduced by quantum channel with a) $n=2$, b) $n=4$, c) $n=3$ and d) $n=5$ for different values of $|c|$ and $|d|$. Parameters $a=0.33$ and $b=0.17$.}
\end{figure}

In case of initial X-state one can obtain matrix $\rho_{X,C}$
\begin{equation}
\rho_{X,C}=(\sigma_y\otimes\sigma_y)\rho_X^*(\sigma_y\otimes\sigma_y)=\rho_X.
\end{equation}
Thus, matrix $R=\rho_{X}^2$ and its eigenvalues are $(A\pm|D|)^2$, $(B\pm|C|)^2$. Utilizing property of considered nonlinear channel $\rho_X\to\Phi(\rho_X)$ that new state density matrix has the structure of X-state density matrix, one can find concurrence for arbitrary integer $n$. The eigenvalues of matrix $R_n=\rho_{X,n}\rho_{X,n,C}$ are $(A_n\pm|D_n|)^2$, $(B_n\pm|C_n|)^2$.

Concurrence takes value $0$ for separable states and greater for entangled states. On figure 6 the concurrence for density matrices $\rho_{X,n}$ for different cases of power $n$ is shown.

\section{Conclusion}
\pst

To resume we point out the main results of our work. We shown that the nonlinear channels acting on initial X-state  of two qubits changes the entanglement characteristics of the state. The entropy of the transformed state is shown to decrease with $n\to\infty$. The entanglement degree expressed by means of concurrence and negativity parameters is shown to increase under influence of the nonlinear channel. The action of nonlinear channels on the other quantum states will be studied in future publication.

\end{document}